\documentclass{article}
\usepackage{spconf,amsmath,graphicx}
\usepackage{amsthm}
\usepackage{multirow}
\usepackage{array}
\usepackage{cite}
\usepackage{amssymb,amsfonts}
\usepackage{textcomp}
\usepackage{color}

\def\0{{\mathbf 0}}
\def\1{{\mathbf 1}}

\def\b{{\mathbf b}}

\def\p{{\mathbf p}}
\def\q{{\mathbf q}}

\def\v{{\mathbf v}}

\def\x{{\mathbf x}}

\def\A{{\mathbf A}}
\def\B{{\mathbf B}}

\def\D{{\mathbf D}}

\def\I{{\mathbf I}}

\def\L{{\mathbf L}}
\def\M{{\mathbf M}}

\def\P{{\mathbf P}}
\def\Q{{\mathbf Q}}

\def\W{{\mathbf W}}

\def\ie{{\textit{i.e.}}}
\def\eg{{\textit{e.g.}}}

\def\cE{{\mathcal E}}

\def\cG{{\mathcal G}}

\def\cO{{\mathcal O}}

\def\cT{{\mathcal T}}

\def\cV{{\mathcal V}}

\def\bTheta{{\boldsymbol \Theta}}

\usepackage[utf8]{inputenc}
\usepackage[english]{babel}

\theoremstyle{plain}

\title{Fast computation of Generalized Eigenvectors for \\ Manifold Graph Embedding}
\name{Fei Chen$^{\dagger}$ \qquad Gene Cheung$^{\star}$ \qquad Xue Zhang$^{\star}$ \thanks{Gene Cheung acknowledges the support of the NSERC grants RGPIN-2019-06271,  RGPAS-2019-00110. Fei Chen acknowledges the support of National Natural Science Foundation of China (61771141).}}
\address{$^{\dagger}$ Fuzhou University, Fuzhou, China ~~~~~~~~~~ $^{\star}$ York University, Toronto, Canada }
\begin{document}
%
\maketitle
\begin{abstract}
Our goal is to efficiently compute low-dimensional latent coordinates for nodes in an input graph---known as graph embedding---for subsequent data processing such as clustering.
Focusing on finite graphs that are interpreted as uniform samples on continuous manifolds (called manifold graphs), we leverage existing fast extreme eigenvector computation algorithms for speedy execution.
We first pose a generalized eigenvalue problem for sparse matrix pair $(\A,\B)$, where $\A = \L - \mu \Q + \epsilon \I$ is a sum of graph Laplacian $\L$ and disconnected two-hop difference matrix $\Q$.
Eigenvector $\v$ minimizing Rayleigh quotient $\frac{\v^{\top} \A \v}{\v^{\top} \v}$ thus minimizes $1$-hop neighbor distances while maximizing distances between disconnected $2$-hop neighbors, preserving graph structure. 
Matrix $\B = \text{diag}(\{\b_i\})$ that defines eigenvector orthogonality is then chosen so that boundary / interior nodes in the sampling domain have the same generalized degrees.
$K$-dimensional latent vectors for the $N$ graph nodes are the first $K$ generalized eigenvectors for $(\A,\B)$, computed in $\cO(N)$ using LOBPCG, where $K \ll N$.
Experiments show that our embedding is among the fastest in the literature, while producing the best clustering performance for manifold graphs. 
\end{abstract}
\vspace{-0.05in}
\section{Introduction}
\label{sec:intro}
\begin{figure}
\centering
\includegraphics[width=2.8in]{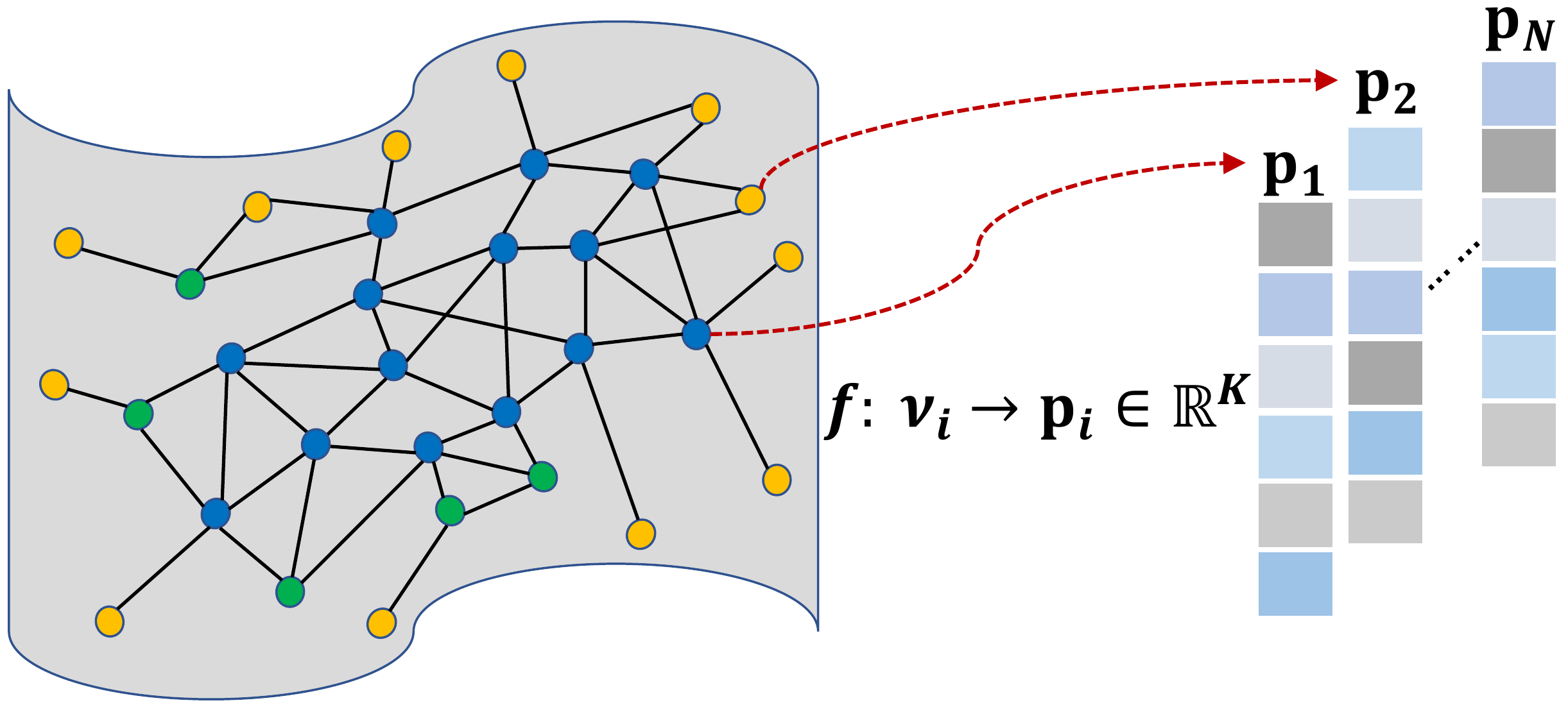}
\vspace{-0.1in}
\caption{\small Example of a finite graph as uniformed sampled points on a smooth continuous manifold. 
Boundary nodes have smaller degrees. 
Nodes are mapped into a low-dimensional latent vector space.}
\label{fig:finite_graph}
\vspace{-0.1in}
\end{figure}

\vspace{-0.05in}
\textit{Graph embedding} is the computation of $K$-dimensional latent space vectors for $N$ nodes in a sparse graph, where typically $K \ll N$ \cite{hamilton17,xu21}.
It means converting an $N \times N$ sparse adjacency matrix to a $N \times K$ dense matrix, resulting in a compact data representation. 
See Fig.\;\ref{fig:finite_graph} for an illustration. 
The smaller representation (while preserving graph structure and nodes' pairwise similarities in the latent vector space) is important for computation and memory requirements when the graph is very large. 
The conversion also enables algorithms and learning models like convolutional neural nets (CNN) \cite{LeCun2015DeepL} designed for vector-space data to process graph-structured data by operating on the new representation directly.

Classical graph embedding methods like \textit{local linear embedding} (LLE) \cite{LLE} and \textit{Laplacian eigenmaps} (LE) \cite{LE} focused exclusively on the large sparse adjacency matrix, and they do not scale well to large graphs.
Recent methods based on conditional probability distributions estimated during different random walks \cite{Deepwalk,node2vec} are more scalable, but still require significant time till convergence.

In an orthogonal development, fast algorithms to compute extreme eigenvectors for real symmetric matrices have been studied extensively the past two decades in numerical linear algebra \cite{golub12}. 
Specifically, \textit{locally optimal block pre-conditioned conjugate gradient} (LOBPCG) \cite{lobpcg} computes $K$ generalized extreme eigen-pairs in $\cO(N)$ for sparse $N \times N$ matrix pair $(\A,\B)$, where $K \ll N$.
Among its wide adoption, LOBPCG was recently used in \textit{graph signal processing} (GSP) \cite{ortega18ieee,cheung18}.
For example, to greedily choose graph samples, at each iteration \cite{anis16} used LOBPCG to compute the first eigenvector of a graph Laplacian sub-matrix and identify the largest magnitude entry.
Another example is \textit{graph metric learning} \cite{yang21}, where LOBPCG was used to compute the first eigenvector $\v \in \mathbb{R}^N$ of a previous matrix solution $\M^t$ at iteration $t$, so that linear constraints can be imposed using scalars $\{s_i\}_{i=1}^N$, where $s_i = 1/v_i$, replacing the \textit{positive semi-definite} (PSD) cone constraint $\M^{t+1} \succeq 0$ for the next solution $\M^{t+1}$. 

Leveraging fast extreme eigenvector algorithms like LOBPCG, in this paper we efficiently compute a graph embedding by formulating a generalized eigenvalue problem with matrix pair $(\A,\B)$. 
Our method is \textit{parameter-free} and targets specifically finite graphs that can be interpreted as uniformly sampled points on low-dimensional continuous manifolds \cite{pang17}---we call them \textit{manifold graphs}. 
Specifically, we first define PSD matrix $\A = \L - \mu \Q + \epsilon \I$, where $\L$ is a graph Laplacian matrix for graph $\cG$, $\Q$ is a difference matrix counting two-hop neighbors that are disconnected, and $\I$ is an identity matrix.
Eigenvector $\v$ minimizing Rayleigh quotient $\frac{\v^{\top} \A \v}{\v^{\top} \v}$ thus minimizes 1-hop neighbor distances, while maximizing distances between disconnected 2-hop neighbors, preserving graph structure.

Second, observing that generalized vector $\v$ for $(\A,\B)$ is also a right eigenvector for asymmetric matrix $\B^{-1} \A$, assuming $\B$ is \textit{positive definite} (PD) and thus invertible, we define $\B = \text{diag}(\{b_i\})$ for strictly positive scalars $\{b_i\}_{i=1}^N$ as follows. 
Local connectivities (node degrees) of a manifold graph $\cG$ should reflect the dimensionality of the underlying manifold \cite{kegl02}. 
The exceptions are boundary nodes of the sampling domain, which have smaller degrees by graph construction;
see Fig.\;\ref{fig:finite_graph} for an illustration. 
As a remedy, we choose scalars $\{b_i\}_{i=1}^N$ so that the \textit{Gershgorin disc} radii \cite{varga04} of matrix $\B^{-1} \A$---row sums of off-diagonal terms in absolute value interpreted as \textit{generalized} node degrees---are the same, while observing the constraint $\prod_{i=1}^N b_i = 1$.

Having defined $(\A,\B)$, a $K$-dimensional graph embedding is generated from the first $K$ generalized eigenvectors, computed using LOBPCG \cite{lobpcg} in linear time $\cO(N)$, assuming $K \ll N$.
Experimental results show that our embedding is among the fastest in the literature, while producing the best clustering performance for manifold graphs. 

\vspace{-0.05in}
\section{Related Work}
\label{sec:related}

\vspace{-0.05in}
\subsection{Graph Embeddings}

As done in \cite{xu21}, existing embedding methods can be classified into three categories: matrix factorization, random walk, and deep learning. 
Matrix factorization-based methods, such as LLE \cite{LLE} and LE \cite{LE}, obtain an embedding by decomposing the large sparse adjacency matrix. 
Complexities of these methods are typically $\cO(N^2)$ and thus are not scalable to large graphs.
Random walk-based methods like Deepwalk \cite{Deepwalk} and node2vec \cite{node2vec} use a random walk process to encode the co-occurrences of nodes to obtain scalable graph embeddings. 
These schemes typically have complexity $\cO(N\log{N})$.
Deep learning approaches, especially autoencoder-based methods \cite{NetWalk} and \textit{graph convolutional network} (GCN)\cite{EGCN}, are also widely studied for graph embedding. 
However, pure deep learning methods require long training time and large memory footprint to store a sizable set of trained parameters.



\subsection{Representation Learning for Graphs}

More generally, one can interpret graph embedding as a task in \textit{representation learning} given an input graph structure.
\cite{hamilton17} formalized this notion in an \textit{encoder-decoder} framework: i) an encoder first maps all graph nodes into $K$-dimensional vectors $\cV \mapsto \mathbb{R}^K$, then ii) a decoder maps every pair of graph embeddings back to a proximity metric, \ie, $\mathbb{R}^K \times \mathbb{R}^K \mapsto \mathbb{R}^+$, that reflects the closeness of two nodes in the original graph.   
A \textit{loss function} can be subsequently defined to measure the quality of the decoder's reconstructed proximities for all node pairs. 
All the aforementioned graph embedding schemes are thus optimizations of encoder parameters to minimize the defined loss function. 

In even more general terms, a node $i$'s proximity information to all other nodes can be directly inputted to an autoencoder for compact representation learning \cite{cao2016deep,wang2016structural}. 
Moreover, each node may be endowed with \textit{attributes} that can influence the notion of node-pair proximity, and GCN can be used to learn representations using a node $i$'s local neighborhood and attributes within \cite{kipf2016semi,schlichtkrull2018modeling}. 
Instead of these recent approaches that progressively increase the generality of the graph embedding problem---resulting in even more complex algorithms---we take the opposite approach of narrowing our focus to embeddings of the important subclass of manifold graphs only. 
This leads to an algorithm based on computation of generalized eigenvectors that is simple, fast and parameter-free. 

\vspace{-0.05in}
\section{Preliminaries}
\label{sec:prelim}
\vspace{-0.05in}
We first provide graph definitions used in our formulation, then review \textit{Gershgorin circle theorem} (GCT) \cite{varga04}.
Finally, we discuss manifold graphs---our graphs of interest.

\subsection{Graph Definitions}

A graph $\cG(\cV,\cE,\W)$ is defined by a set of $N$ nodes $\cV = \{1, \ldots, N\}$, edges $\cE = \{(i,j)\}$, and an \textit{adjacency matrix} $\W$. 
$W_{i,j} \in \mathbb{R}^+$ is the positive edge weight if $(i,j) \in \cE$, and $W_{i,j} = 0$ otherwise. 
\textit{Degree matrix} $\D$ is a diagonal matrix with entries $D_{i,i} = \sum_{j} W_{i,j}, \forall i$. 
A \textit{positive semi-definite} (PSD) \textit{combinatorial graph Laplacian matrix} $\L$ for a positive graph is defined as $\L \triangleq \D - \W$ \cite{cheung18}. 

A graph signal $\x \in \mathbb{R}^N$ is smooth with respect to (wrt) graph $\cG$ if its \textit{graph Laplacian regularizer} (GLR), $\x^{\top} \L \x$, is small~\cite{pang17}:
\begin{align}
\x^{\top} \L \x = \sum_{(i,j) \in \cE} W_{i,j} (x_i - x_j)^2 .
\end{align}
GLR is commonly used to regularize ill-posed inverse problems such as denoising or dequantization \cite{pang17,liu17}.

\subsection{Gershgorin Circle Theorem}

Given a real symmetric square matrix $\M \in \mathbb{R}^{N \times N}$, corresponding to each row $i$ is a \textit{Gershgorin disc} $i$ with center $c_i \triangleq M_{i,i}$ and radius $r_i \triangleq \sum_{j \neq i} |M_{i,j}|$. 
By GCT \cite{varga04}, each eigenvalue $\lambda$ of $\M$ resides in one (or more) Gershgorin disc, \ie, $\exists i$ such that 
\begin{align}
c_i - r_i \leq \lambda \leq c_i + r_i .
\label{eq:GCT}
\end{align}
A corollary is that the smallest Gershgorin disc left-end $\lambda^-_{\min}(\M)$ is a lower bound of the smallest eigenvalue $\lambda_{\min}(\M)$ of $\M$, \ie,
\begin{align}
\lambda^-_{\min}(\M) \triangleq \min_i c_i - r_i \leq \lambda_{\min}(\M) .
\label{eq:GCT}
\end{align}

\subsection{Manifold Graphs} 
\label{subsec:manifold}

We target our embedding specifically for \textit{manifold graphs}, which are finite graphs interpreted as uniformly sampled points on smooth continuous manifolds.
This is the commonly held \textit{manifold hypothesis} \cite{carey2017graph}: though points in a dataset are observed in a high-dimensional input space, they intrinsically reside in a lower-dimensional manifold space upon an appropriate transformation.
There exist numerous graph construction algorithms \cite{carreira2005proximity, liu2011mixture,carey2017graph} that select node samples closely approximating this hypothesized manifold. 
To evaluate quality of a constructed graph, \cite{carey2017graph} proposed several metrics; one example is \textit{betweenness centrality}, which measures how often a node $i$ appears in a shortest path between two nodes in the graph. 
Mathematically, it is defined as
\begin{align}
C_B(i)=\sum_{s,t\neq i} \frac{\sigma_{st}(i)}{\sigma_{st}}
\end{align}
where $\sigma_{st}$ is the number of shortest paths from node $s$ to node $t$ and $\sigma_{st}(i)$ is the number of those paths that pass through node $i$. 
Given a graph composed of nodes uniformly sampled from a smooth continuous manifold, the betweenness centrality of nodes should be similar, \ie, all nodes are equally likely to appear in a given shortest path.
Thus, we employ the \textit{variance of betweenness centrality} (VBC) as our metric to evaluate the quality of a manifold graph; only qualified manifold graphs are inputted to our proposed algorithm. 
As shown in Table.\;\ref{tab:vbc}, the first four graphs with smaller VBCs are considered qualified manifold graphs to include in our experiments in Section\;\ref{sec:results}. 

\begin{table}   
\begin{center}   
\footnotesize
\caption{VBCs $(\times10^{5})$ of graphs}
\label{tab:vbc} 
\begin{tabular}{|c|c|c|c|c|c|c|} 
\hline   Jaffe & AT\&T & Karate & Football & Citeseer & Cora & AUS \\   
\hline    2.47 & 8.57 & 0.02 & 0.01 & 6.6e+03 & 7.0e+03 & 1.1e+03 \\ 
\hline   
\end{tabular}   
\end{center}
\vspace{-0.5cm}
\end{table}


\vspace{-0.05in}
\section{Computing Embeddings}
\label{sec:embedding}


\vspace{-0.05in}
\subsection{Defining Objective}

We first define $\P \in \mathbb{R}^{N \times K}$, where the $i$-th row of $\P$ contains the $K$-dimensional latent vector $\p_i \in \mathbb{R}^K$ for node $i \in \cV$.
For notation convenience, we define also $\q_k$ as the $k$-th column of $\P$---the $k$-th coordinate of all $N$ nodes. 
To minimize the latent space distances between connected $1$-hop neighbors $(i,j) \in \cE$ in original graph $\cG$, we first minimize the GLR \cite{pang17}:
\begin{align}
\min_{\P \,|\, \P^{\top} \P = \I} \text{tr} \left( \P^{\top} \L \P \right) 
&= \sum_{k=1}^K \q_k^{\top} \L \q_k 
\label{eq:obj1} \\
&= \sum_{k=1}^K \sum_{(i,j) \in \cE} w_{i,j} (q_{k,i} - q_{k,j})^2 
\nonumber 
\end{align}
where $q_{k,i}$ is the $k$-th latent coordinate of node $i$.
Like LLE \cite{LLE}, orthogonality condition $\P^{\top} \P = \I$ is added to ensure $\q_i^{\top} \q_j = \delta_{i-j}$. 
Minimizing \eqref{eq:obj1} would minimize the squared Euclidean distance $\|\p_i - \p_j\|^2_2$ between connected node pair $(i,j)$ in the latent space.
This objective thus preserves the \textit{first-order proximity} of the original graph structure \cite{xu21}.

\begin{figure}
\centering
\begin{tabular}{cc}
\includegraphics[width=1.45in]{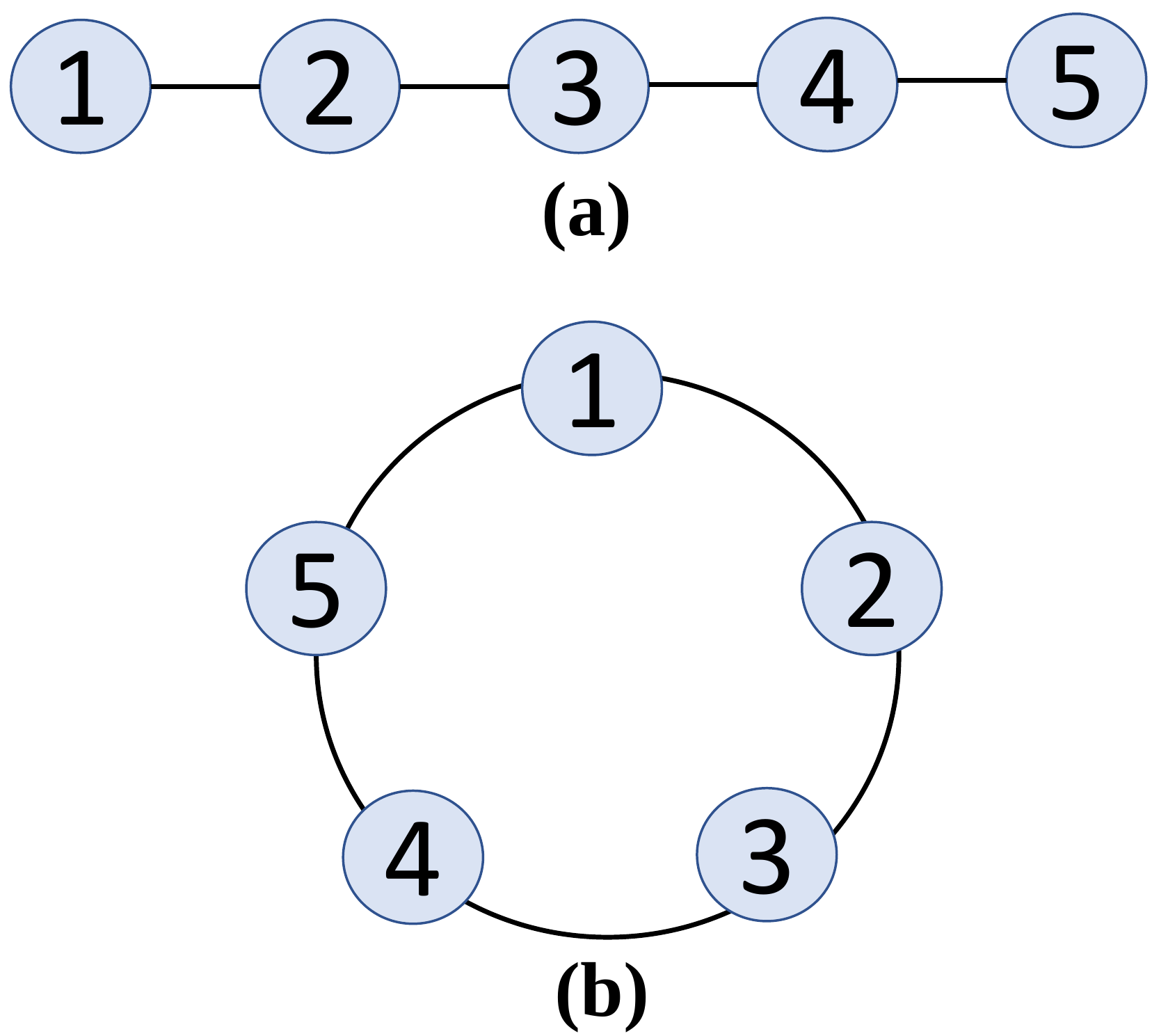}&\includegraphics[width=1.7in]{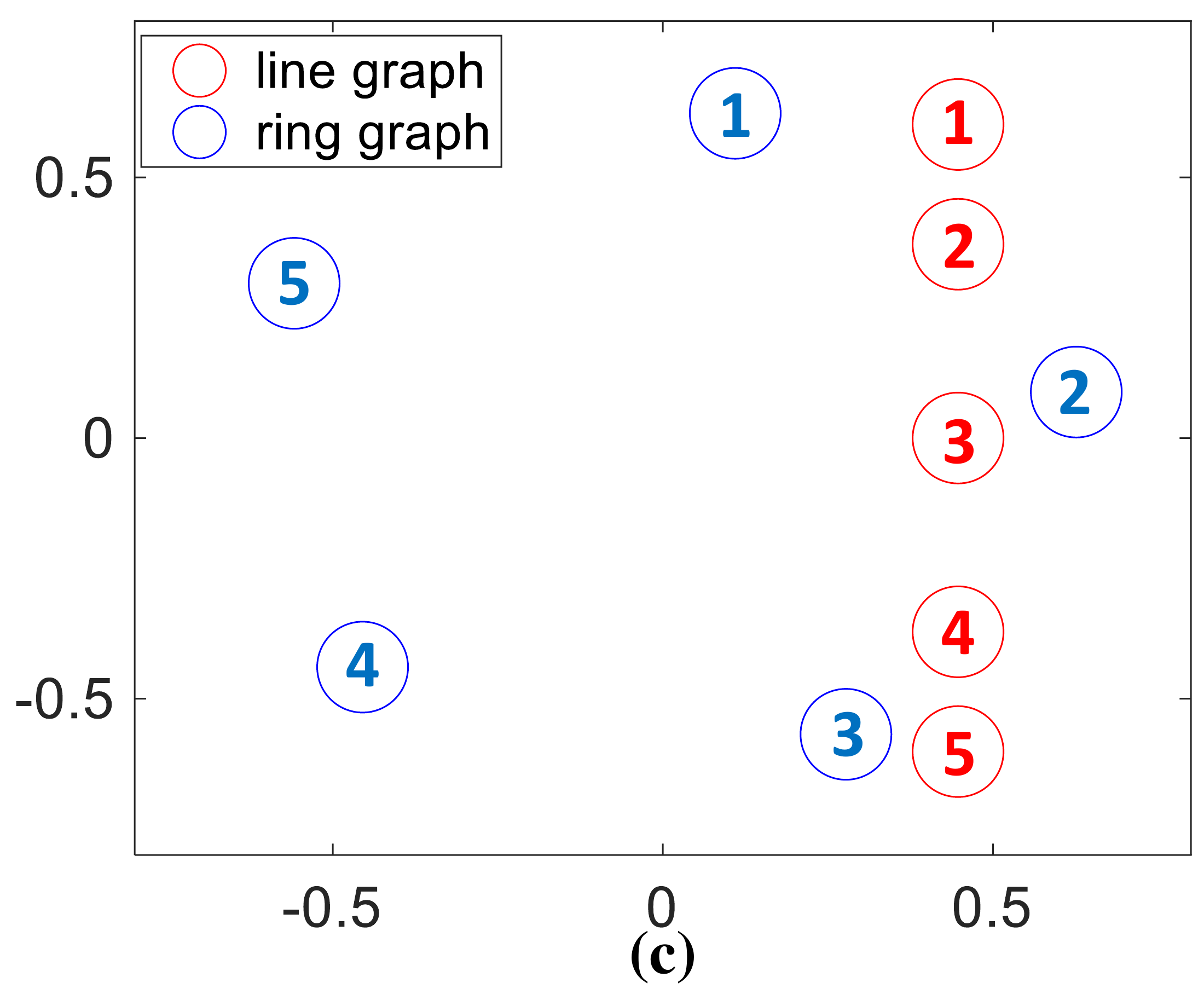}\\
\vspace{-0.25in}
\end{tabular}
\caption{\small Illustration of (a) a $5$-node line graph, (b) a $5$-node ring graph, where all nodes have the same degrees, and (c)  the first eigenvectors (\ie, 2D latent space vectors) of $\A$'s of (a) and (b).}
\label{fig:4node_graph}
\vspace{-0.1in}
\end{figure}

\vspace{-0.05in}
\subsubsection{$2$-hop Neighbor Regularization}

However, objective \eqref{eq:obj1} is not sufficient---it does not consider \textit{second-order proximity} of the original graph $\cG$. 
Consider the simple $5$-node line graph example in Fig.\;\ref{fig:4node_graph}(a). 
Just requiring each connected node pair to be located in close proximity is not sufficient to uniquely induce a straight line solution (and thus in lowest dimensional latent space). 
For example, a zigzag line in 2D latent space is also possible.

Thus, we regularize the objective \eqref{eq:obj1} using our second graph assumption: 
\textit{sparsity of the input manifold graph connectivity is determined based on point-to-point distance on the underlying manifold}. 
In other words, if $(i,j) \in \cE$ but $(i,l) \not\in \cE$, then manifold distance $d_{i,j}$ between $(i,j)$ must be smaller than distance $d_{i,l}$ between $(i,l)$, or $d_{i,j} < d_{i,l}$. 

Based on this assumption, we define our regularizer $g(\P)$ as follows.
Denote by $\cT_i$ the \textit{two-hop neighbor} node set from node $i$; \ie, node $j \in \cT_i$ is reachable in two hops from $i$, but $(i,j) \not\in \cE$.
The aggregate distance between each node $i$ and its 2-hop neighbors in $\cT_i$ is
$\sum_{i \in \cV} \sum_{j \in \cT_i} \|\p_i - \p_j \|^2_2$.

We write this aggregate distance in matrix form.
For each $\cT_i$, we first define matrix $\bTheta_i \in \mathbb{R}^{N \times N}$ with entries

\vspace{-0.1in}
\begin{small}
\begin{align}
\Theta_{i,m,n} = \left\{ \begin{array}{ll}
\frac{1}{T_i} & \mbox{if}~ m = n = i ~~~\mbox{or}~~~ m = n \in \cT_i \\
-\frac{1}{T_i} & \mbox{if}~ m = i, n \in \cT_i ~~~\mbox{or}~~~ m \in \cT_i, n = i \\
0 & \mbox{o.w.}
\end{array} \right. 
\label{eq:Q}
\end{align}
\end{small}\noindent
where $T_i = |\cT_i|$ is the number of disconnected 2-hop neighbors.
We then define $\Q = \sum_{i \in \cV} \bTheta_i$. 
Finally, we define the regularizer as $g(\P) = - \mu \, \text{tr}(\P^{\top} \Q \P) + \epsilon \I$, where $\I$ is the identity matrix.
Parameters $\mu, \epsilon > 0$ and are chosen to ensure matrix PSDness (to be discussed).
The optimization becomes
\begin{align}
\min_{\P} &~ \text{tr}(\P^{\top} \L \P) - \mu \, \text{tr}(\P^{\top} \Q \P) + \epsilon \I 
\nonumber \\ 
&= \text{tr} (\P^{\top} \underbrace{\left( \L - \mu \Q + \epsilon \I \right)}_{\A} \P ) .
\label{eq:obj2}
\end{align}
Note that objective \eqref{eq:obj2} remains quadratic in variable $\P$. 


\subsection{Choosing Weight Parameter $\mu$}

As a quadratic minimization problem \eqref{eq:obj2}, it is desirable for $\A = \L - \mu \Q + \epsilon \I$ to be PSD so that the objective is lower-bounded, \ie,  $\q^{\top} \A \q \geq 0, \forall \q \in \mathbb{R}^N$. 
We set $\epsilon = \lambda^{(2)}_{\min}(\Q)$ to be the \textit{second} smallest eigenvalue---the Fiedler number---of $\Q$ (Laplacian has $\lambda^{(1)}_{\min}(\Q) = 0$); larger $\lambda^{(2)}_{\min}(\Q)$ means more disconnected 2-hop neighbors, and a larger $\mu$ is desired. 
We compute $\mu > 0$ so that $\A$ is guaranteed to be PSD via GCT \cite{varga04}.
Specifically, we compute $\mu$ such that left-ends of all Gershgorin discs $i$ corresponding to rows of $\A$ (disc center $A_{i,i}$ minus radius $\sum_{j \neq i} |A_{i,j}|$) are at least 0, \ie,
\begin{align}
L_{i,i} - \mu Q_{i,i} + \epsilon - \sum_{j | j \neq i} \left| L_{i,j} - \mu Q_{i,j} \right| & \geq 0, ~~\forall i .
\label{eq:discLE1}
\end{align}

Note that $L_{i,j} = -W_{i,j} \leq 0$, and $Q_{i,j} \leq 0$.
Note further that node $j$ cannot both be a $1$-hop neighbor to $i$ and a disconnected $2$-hop neighbor at the same time, and hence either $L_{i,j}=0$ or $Q_{i,j}=0$. 
Thus, we can remove the absolute value operator as
\begin{align}
L_{i,i} - \mu Q_{i,i} + \epsilon - \sum_{j | j \neq i} \left( -L_{i,j} - \mu Q_{i,j} \right) &\geq 0 .
\label{eq:discLE2}
\end{align}
We set the equation to equality and solve for $\mu_i$ for row $i$, \ie, 

\vspace{-0.15in}
\begin{small}
\begin{align}
\mu_i = \frac{L_{i,i} + \sum_{j|j\neq i} L_{i,j} + \epsilon}{Q_{ii} - \sum_{j|j\neq i} Q_{i,j}} 
= \frac{\epsilon}{Q_{ii} - \sum_{j|j\neq i} Q_{i,j}} .
\label{eq:mu}
\end{align}
\end{small}\noindent
where $L_{i,i} = - \sum_{j\neq i} L_{i,j}$.
Finally, we use the smallest non-negative $\mu = \min_{i} \mu_i$ for \eqref{eq:obj2} to ensure all disc left-ends are at least $0$, as required in \eqref{eq:discLE1}.

\subsection{Defining Orthogonality Condition}


Instead of orthogonality condition $\q_i^{\top} \q_j = \delta_{i-j}$, we generalize the condition to $\q_i^{\top} \B \q_j = \delta_{i-j}$ for a chosen \textit{positive definite} (PD) matrix $\B$, which also implies the desirable $\q_i \neq \q_j$ for $i \neq j$.  
For simplicity, we choose $\B = \text{diag}(\{b_i\})$, where $b_i > 0, \forall i$, are scalars for the $N$ nodes. 
The constraint on variable $\P$ is now $\P \B \P^{\top} = \I$, which together with objective \eqref{eq:obj2} means computing the first $K$ \textit{generalized eigen-pair} for matrix pair $(\A, \B)$. 
We use LOBPCG \cite{lobpcg} for this task, running in $\cO(N)$ for sparse matrices, assuming $K \ll N$. 

For a given generalized eigen-pair $(\lambda_, \q)$ for matrix pair $(\A,\B)$ where $\B$ is PD and hence invertible, we can write
\begin{align}
\A \q &= \lambda \B \q \\
\B^{-1} \A \q &= \lambda \q .
\end{align}
Thus, $\q$ is also a \textit{right} eigenvector for asymmetric matrix $\B^{-1} \A$.
We determine scalars $\{b_i\}_{i=1}^N$ from this right eigenvector perspective.

We first observe that the contribution of node $i$ in the numerator $\q^{\top} \A \q$ of the Rayleigh quotient for matrix $\A$ is
\begin{align}
\sum_{j \in \cE_i} w_{i,j} (q_i - q_j)^2 - \mu \sum_{j \in \cT_i} (q_i - q_j)^2, ~~ \forall i \in \cV
\label{eq:contri}
\end{align}
where $\cE_i = \{j \,|\, (i,j) \in \cE\}$ is the set of $1$-hop neighbors of node $i$. 
In general, boundary nodes in the sampling domain have smaller degrees than interior nodes, as illustrated in Fig.\;\ref{fig:finite_graph}, and thus smaller contributions to the Rayleigh quotient.
As shown in the $5$-node line graph in Fig.\;\ref{fig:4node_graph}, because degrees of boundary nodes do not reflect the dimensionality of the underlying manifold, they create problems when deriving latent coordinates from computed eigenvectors; connected node-pairs in a line graph are not evenly spaced (\eg, pair $(1,2)$ are closer than $(2,3)$), while a ring graph with no boundary nodes has no such problem.

We design scalars $\{b_i\}_{i=1}^N$ to remedy this problem. 
We first define \textit{generalized node degree} for node $i$ as the Gershgorin disc radius of row $i$ of $\B^{-1} \A$. 
We choose $\{b_i\}_{i=1}^N$ so that the generalized node degrees of all nodes are the same, while $\prod_{i=1}^N b_i = 1$.
Specifically, $r_1/b_1=r_2/b_2=\cdots=r_N/b_N$, where $r_i$ is the Gershgorin disc radius of row $i$ of $\A$. Then we have $b_1={r_1}/{(\prod_{i=1}^N{r_i})^{1/N}}$ and $b_i=r_ib_1/r_1$.


\vspace{-0.05in}
\section{Experiments}
\label{sec:results}
\vspace{-0.05in}
\subsection{Experimental Setup}

We conducted extensive experiments to test our embedding method. 
We compared it with representative state-of-art embedding methods: i) matrrix factorization-based (LLE\cite{LLE}, LE \cite{LE}), ii) random walk-based (DeepWalk\cite{Deepwalk}, node2vec\cite{node2vec}), and deep learning-based (NetWalk\cite{NetWalk}).
DeepWalk and NetWalk were executed with default hyperparameters. 
For node2vec, we used the following parameters: window size 10, walk length 20, and walk number 200. 
All our experiments were run in the Matlab2015b environment on a laptop with Intel Core i5-8365U CPU of 1.60GHz.

We first tested k-nearest neighbor (kNN) graphs. 
Since face datasets have a known low-dimensional manifold structure, two kNN graphs were constructed from datasets JAFFE  \cite{JAFFE} and AT\&T \cite{ATT} based on Euclidean distances between facial images.
For a graph with unknown construction, we use VBC discussed in Section\;\ref{subsec:manifold} to evaluate its quality as a manifold graph.  
As shown in Table\;\ref{tab:vbc}, Karate \cite{Karate} and Football\cite{Girvan2002football} (social network datasets)
have relatively small VBCs and thus are better manifold graphs than JAFFE and AT\&T.

The number of clusters $C$ in each graph is the number of known clusters in the dataset, then kmeans and Gaussian mixture model (GMM) are used for unsupervised graph clustering. 
To evaluate clustering accuracy, we used four criteria: Rand Index (RI), Precision, Purity, and Normalized Mutual Information (NMI) \cite{Alok2013DevelopmentOA}. 
These values range from 0 to 1, and a higher value indicates better clustering.

\begin{table}   
\begin{center}   
\caption{Description of datasets}  
\label{tab:dataset} 
\begin{footnotesize}
\begin{tabular}{|c|c|c|c|c|}   
\hline   Dataset & $N$ & $|\cE|$ & $K$& $C$  \\   
\hline   JAFFE & 200 & 1822  & 2 & 10  \\ 
\hline   AT\&T & 400 & 2054  & 2& 40\\  
\hline   Karate &  34 & 154 & 2 &2  \\   
\hline   Football& 105  & 613 &2 & 12  \\  
\hline   
\end{tabular}   
\end{footnotesize}
\end{center}   
\end{table}

\subsection{Experimental Results}

\begin{figure}
\centering
\includegraphics[width=3.2in]{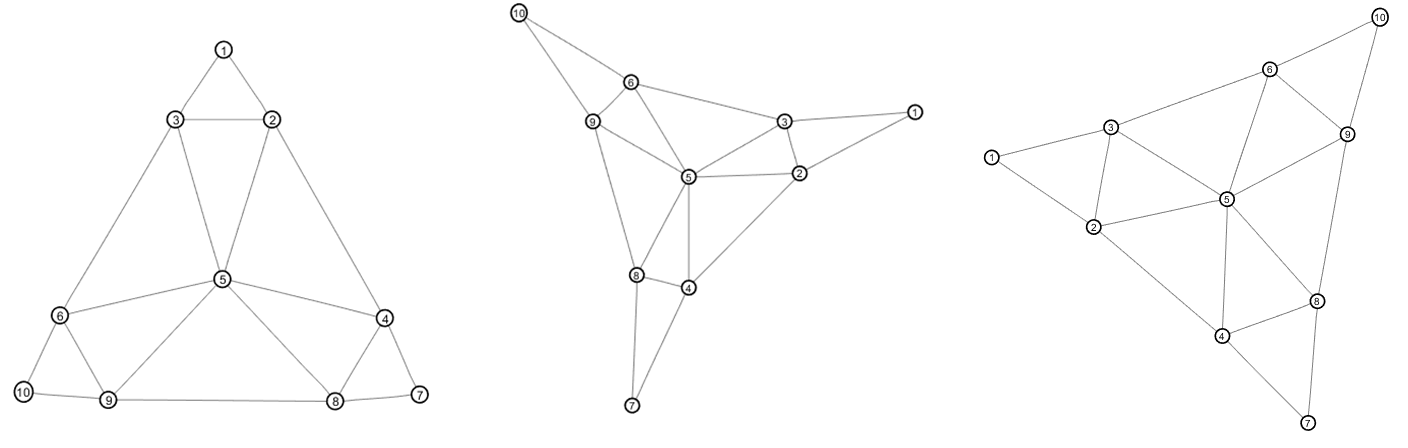}
\vspace{-0.15in}
\caption{\small Triangle mesh. Left: LLE, middle: LE, right: proposed.}
\label{fig:mesh}
\vspace{-0.1in}
\end{figure}

We first visualize the resulting embeddings from our method, LLE and LE for a 10-node triangle mesh graph, containing 18 equal edges. 
Fig.\;\ref{fig:mesh} shows the embeddings in 2D space. 
Since corner nodes cannot be represented as linear combinations of respective neighbor nodes, there were obvious shape distortions at the boundaries by LLE. 
LE described local one-hop connections using the Laplacian matrix, but it ignored disconnected two-hop neighbors when computing latent vectors. 
In contrast, our method achieved roughly equal node spacings using a disconnected 2-hop neighbor regularization term.

Table\;\ref{tab:numerical} lists the quantitative comparison results against all competing methods.
The best results of each criterion are in boldface. 
Our method achieved very competitive performance on two kNN graph datasets and two social network datasets.  
We observe that the performance of LE was the worst. 
Thus, the disconnected two-hop neighbor regularization helped preserved original graph structure. 
LLE achieved good performance on kNN graphs, but it became unstable and did not work well for Karate. 

For datasets with fewer classes, Deepwalk and node2vec were competitive and worked better than LLE and LE, especially when the graph had many edges between nodes from different clusters. 
node2vec performed biased random walks on the graph and embedded nodes appearing together in the embedding space. 
We observe that node2vec outperformed LLE and LE on three datasets. 
The random walk-based methods had slower convergence with approximate inference techniques. 
The time complexity of random walk is $\cO(N\log{N})$. 
See Table\;\ref{tab:complex} for complexity comparison of different methods. 
NetWalk employed autoencoder to preserve graph proximities, but the training process required a large number of iterations till convergence. 
In contrast, our method was fast using LOBPCG \cite{lobpcg} and is by design entirely parameter-free.

\begin{table}   
\begin{center}   
\caption{Clustering performance of five embedding methods in terms of four criteria. 
Using obtained latent vectors, we employed kmeans and GMM for unsupervised clustering. 
Each criterion is computed using the average results of kmeans and GMM.}

\label{tab:numerical} 
\begin{scriptsize}
\begin{tabular}{|c|c|c|c|c|c|}   
\hline   Dataset& Method & RI & Precision & Purity & NMI\\   
\hline   \multirow{6}*{JAFFE}&LE &0.816	&0.666&	0.768&	0.620 \\ 
\cline{2-6}   &LLE &0.878&	0.758&	0.840&	0.718 \\  
\cline{2-6}    &Deepwalk & 0.908&	0.731&	0.825&	0.753 \\   
\cline{2-6}     &node2vec&0.904	&0.697&	0.800&	0.743 \\  
\cline{2-6}     &NetWalk& 0.912&\textbf{0.810}&	\textbf{0.885}&	0.787 \\   
\cline{2-6}     &proposed& \textbf{0.938}&0.807&	0.875&	\textbf{0.829} \\ 
\hline    \multirow{6}*{AT\&T}&LE & 0.895&	0.377&	0.544&	0.552 \\ 
\cline{2-6}   &LLE &0.925&	0.509&	0.654&	0.647  \\  
\cline{2-6}   &Deepwalk & 0.939&	0.431&	0.583&	0.631 \\   
\cline{2-6}   &node2vec&0.948&	0.405&	0.576&	0.647 \\  
\cline{2-6}   &NetWalk& 0.934&	0.502&	0.648&	0.670 \\   
\cline{2-6}   &proposed&\textbf{0.959}&	\textbf{0.516}&	\textbf{0.661}&	\textbf{0.722} \\ 
\hline   \multirow{6}*{Karate}&LE & 0.529&	0.779&	0.853&	0.207 \\
\cline{2-6}   & LLE & 0.802&	0.842&	0.882&	0.624 \\  
\cline{2-6}   &  Deepwalk & 0.886&	0.890&	0.941&	0.732\\   
\cline{2-6}   &  node2vec& 0.886&  0.890& 0.941& 0.732\\  
\cline{2-6}   &   NetWalk& \textbf{0.941}&	\textbf{0.941}&\textbf{	0.971}&	\textbf{0.837}\\   
\cline{2-6}   &   proposed& \textbf{0.941}&	\textbf{0.941}&\textbf{	0.971}&	\textbf{0.837}\\     
\hline   \multirow{6}*{Football}&LE & 0.916&	0.670&	0.748&	0.729 \\ 
\cline{2-6}   &   LLE &0.916&	0.664&	0.734&	0.719 \\  
\cline{2-6}   &  Deepwalk &  0.851&0.460&0.596&0.551\\   
\cline{2-6}   &   node2vec&0.886&0.493&0.622&0.609\\  
\cline{2-6}   &   NetWalk& 0.849&0.518&0.648&0.575 \\   
\cline{2-6}   &   proposed&\textbf{0.930}&\textbf{0.684}&\textbf{0.761}&\textbf{0.752}\\   
\hline   
\end{tabular}   
\end{scriptsize}
\end{center}   
\end{table}

\begin{table}   
\begin{center} 
\caption{Time complexity of embedding mehods}  
\label{tab:complex} 
\begin{tiny}
\begin{tabular}{|c|c|c|c|c|c|}   
\hline   LLE & LE & DeepWalk & note2vec & NetWalk & Proposed \\   
\hline  $\cO(N^2)$ &$\cO(N^2)$  & $\cO(N\log{N})$   &$\cO(N\log{N})$ & $\cO(N+|\cE|)$ & $\cO(N)$ \\ 
\hline   
\end{tabular}   
\end{tiny}
\end{center}   
\end{table}

\vspace{-0.05in}
\section{Conclusion}
\label{sec:conclude}
\vspace{-0.05in}
Graph embedding---computing a compact $K$-dimensional vector representation for each node while preserving pairwise proximity and graph structure---is an important tool for geometric data processing.
Leveraging existing fast extreme eigenvector algorithms, we presented a fast parameter-free method based on a new generalized eigenvalue formulation with sparse matrix pair $(\A,\B)$. 
We chose $\A$ to minimize $1$-hop neighbor distances while maximizing distances between disconnected $2$-hop neighbors.
We chose $\B$ so that all nodes have the same generalized degree reflecting the intrinsic low dimension of the underlying manifold. 
The first $K$ generalized eigenvectors were computed using LOBPCG in $\cO(N)$, where $K \ll N$. 
Experiments show that our method was fast and produced the best clustering performance.


\vfill\pagebreak

\begin{small}
\bibliographystyle{IEEEbib}
\bibliography{ref2}
\end{small}

\end{document}